# CYBER THREATS IN SOCIAL NETWORKING WEBSITES

Wajeb Gharibi[1] and Maha Shaabi[2]
College of Computer Science & Information Systems
Jazan University, Kingdom of Saudi Arabia
[1]gharibi@jazanu.edu.sa    [2]mshaabi@jazanu.edu.sa

*ABSTRACT*

*A social network is a social structure made up of individuals or organizations called nodes, which are connected by one or more specific types of interdependency, such as friendship, common interest, and exchange of finance, relationships of beliefs, knowledge or prestige. A cyber threat can be both unintentional and intentional, targeted or non targeted, and it can come from a variety of sources, including foreign nations engaged in espionage and information warfare, criminals, hackers, virus writers, disgruntled employees and contractors working within an organization. Social networking sites are not only to communicate or interact with other people globally, but also one effective way for business promotion. In this paper, we investigate and study the cyber threats in social networking websites. We go through the amassing history of online social websites, classify their types and also discuss the cyber threats, suggest the anti-threats strategies and visualize the future trends of such hoppy popular websites.*

## 1. INTRODUCTION

Nowadays, millions of internet users regularly visit thousands of social website to keep linking with their friends, share their thoughts, photos, videos and discuss even about their daily-life. Social networks can be traced back to the first email which was sent in 1971 where two computers were sitting right next to each other. In 1987 Bulletin Board System exchanged data over phone lines with other users and lately in the same year the first copies of early web browsers were distributed through Usenet. Geocities was the first social website founded in 1994. Theglobe.com launched in 1995 and gave people the ability of interacting with others, personalize and publish their files on the Internet. In 1997, the America on Line (AOL) Instant Messenger was lunched. In 2002, Friendster was lunched and within three months more than 3 million users were using it. In 2003, MySpace was lunched and in the following years many other social networking sites were lunched such as Face book in 2004, Twitter in 2006 etc. (See Figure 1) [1].

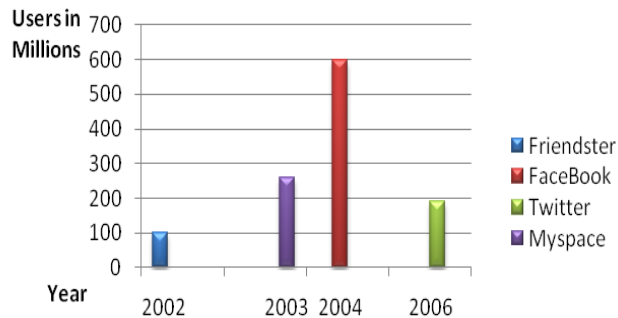

Figure 1. Social Networks compared by users

There are so many social networking sites and social media sites that there is even search engine for them [2]. Further, there are specialized websites which allow users to create their social networking sites such as Ning and KickAppls [3].

These social websites have had positive and negative impacts; so many people waste most of their time on using these websites, which results in losing their jobs or colleges or even their natural social lives and families! Many others post copyrighted materials without authorizations, or pornographic or illegal contents. Some of the users, smart-users, use social networking websites in a very positive way; as happen now in the Spring of Arab World!

Most social networks have members create and manage their personal profiles, post different types of files, provide facilities for members to automatically discover connection candidates with existing members and provide more sophisticated features that designed to have users spend long time on these sites. Moreover, many software developers are working over new advanced applications for such media and networking websites.

Commonly, users make many risks and mistakes when using social networks services such as using unauthorized programs, misuse of corporate computers, unauthorized physical and network access, misuse of passwords and transfer sensitive information between their work and personal computers when working at homes. However, the excessive trust between users of social networks can be used to perpetrate a variety of attacks and data leakage [4].

Due the fact that the number of social networks users is increasing day by day (Ref. Figure 2), the number of attacks carried out by hackers to steal personal information is also raised. Hacked information can be used for many purposes such as sending unauthorized messages (spam), stealing money from victim's accounts etc. The purpose of this paper is to study and analyze the current threats of social network and develop measures to protect the identity in cyberspace i.e., security of personal information and identity in social networks are studied [5-11].

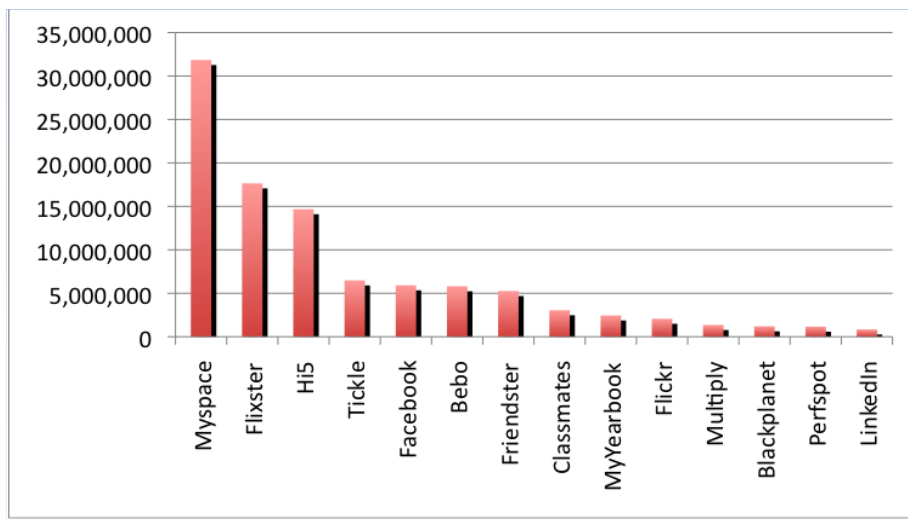

Figure 2. Total number of social networks users (Rapleaf's data)

The Internet today, unfortunately, offers to the cybercriminals many chances to hack accounts on social network sites and the number of malicious programs that target the social web sites is very huge. (Ref: Figure 3)

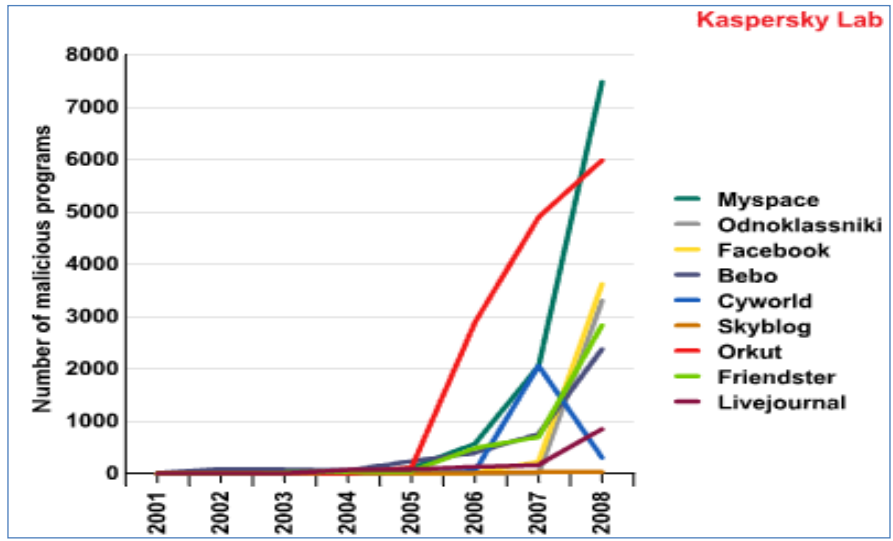

Figure 3. Number of malicious programs targeting popular social networking sites

The rest of the paper is organized as follows. Section 2 summarizes the related works on privacy and threats of social networking websites, In Section 3, we present the categorized types of social networks. We discuss and analysis the cyber threats in social websites in Section 4. In Section 5, it has been recommended the anti-threats strategies. In Section 6, the future trends of social networks have been analyzed. In Section 7, we reveal the risks prevention and threats vulnerabilities. Finally, we ended our paper with the conclusion at Section 8.

## 2. RELATED WORKS

The popularity of the term social networking web sites has been increased since 1997, and millions of people now are using social networking web sites to communicate with their friends, perform business and many other usages according to the interest of the users.

The interest of social networking web sites has been increased and many research papers have been published. Some of them discussed the security issues of social networking, analyzing the privacy and the risks that threat the online social networking web sites.

The article [7] identifies the security behavior and attitudes for social network users from different demography groups and assess how these behaviors map against privacy vulnerabilities inherent in social networking applications.

In the article [8], the researcher highlights the commercial and social benefits of safe and well-informed use of social networking web sites. It emphasizes the most important threats of the users and illustrates the fundamental factors behind those threats. Moreover, it presents the policy and technical recommendations to improve privacy and security without compromising the benefits of the information sharing through social networking web sites.

In [11], author addresses security issues, network and security managers, which often turn to network policy management services such as firewall, intrusion, perfusion system, antivirus and data lose. It addresses security, framework to protect corporation information against the threats related to social networking web sites.

Also many other scientific research papers have been published [12, 13] where the new technology and strategies were discussed related to the privacy and security issues of social networking websites.

# 3. TAXONOMY OF SOCIAL NETWORKS

Generally, a social network is a social structure made up of individuals or organizations which are connected by one or more specific types of interdependency, such as friendship, common interest, and exchange of finance, relationships of beliefs, knowledge or prestige. Social networks can also be defined as those websites that enable people to form online communications and exchange all types of data. It includes the following.

First, Social networking sites such as MySpace, Face book, Windows Live Spaces, Habbo, etc. and the second Social media sites such as You tube, Flicker, Digg ,Metacafe, etc. Table 1 illustrates the social websites according to continent and regions.

Table 1. Social websites according to Continent and Region

| Continent/region | Dominant social websites |
|---|---|
| Africa | Hi5, Facebook |
| America (North) | MySpace, Facebook, Youtube, Flicker, Netlog |
| America (Central &South) | Orkut, Hi5, Facebook |
| Asia | Friendster, Orkut, Xianonei, Xing, Hi5, Youtube, Mixi |
| Europe | Badoo, Bedo, Hi5, Facebook, Xing, Skyrock, Ployaheod, Odnoklassniki.ru.V Kontakte |
| Middle East | Facebook |
| Pacific Island | Bedo |

In table 2, the top five popularity trafficked social media sites:

Table 2. Top five popularity trafficked social media sites

| Site Name | Primary Shared Media |
|---|---|
| YouTube | Videos |
| Flicker | Images |
| Digg | Book marks |
| Metacafe | Videos |
| Stumbleupon | Cool Contents |

Moreover, Youtube is the third most visited Web Site after Yahoo and Google but flicker is the $39^{th}$ most visited web site [5].

# 4. CYBER THREATS IN SOCIAL NETWORKING WEBSITES

Lately, social networks attract thousands of users who represent potential victims to attackers from the following types (Ref: Figure 4) [6, 7].

First Phishers and spammers who use social networks for sending fraudulent messages to victims "friend", Cybercriminals and fraudsters who use the social networks for capturing users data then carrying out their social-engineering attacks and Terrorist groups and sexual predators who create online communities for spreading their thoughts, propaganda, views and conducting recruitment.

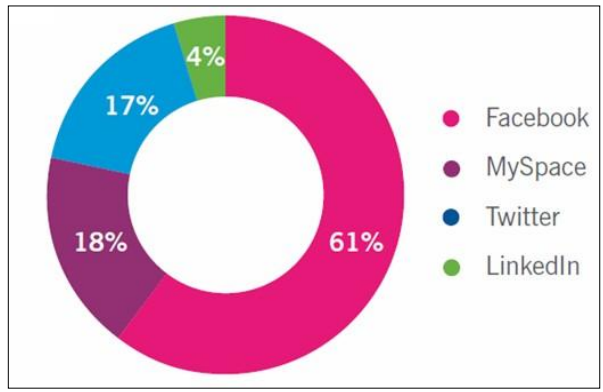

Figure 4. Threats percentage-pose on social networks (Sophos 2010 Security Threat Report)

Cyber threats that might the users face can be categorized into two categories.

### 4.1.1 Privacy Related Threats

Privacy concerns demand that user profiles never publish and distribute information over the web. Variety of information on personal home pages may contain very sensitive data such as birth dates, home addresses, and personal mobile numbers and so on. This information can be used by hackers who use social engineering techniques to get benefits of such sensitive information and steal money.

### 4.1.2. Traditional Networks Threats

Generally, there are two types of security issues: One is the security of people. Another is the security of the computers people use and data they store in their systems. Since social networks have enormous numbers of users and store enormous amount of data, they are natural targets spammers, phishing and malicious attacks. Moreover, online social attacks include identity theft, defamation, stalking, injures to personal dignity and cyber bulling. Hackers create false profiles and mimic personalities or brands, or to slander a known individual within a network of friends.

## 5. ANTI THREATS STRATEGIES

In this section we present the different types of cyber threats in social networks and found the most of threats happens due to the factors which are listed as below:

a) Most of the users are not concern with the importance of the personal information disclosure and thus they are under the risk of over disclosure and privacy invasions.

b) Users, who are aware of the threats, unfortunately choose inappropriate privacy setting and manage privacy preference properly.

c) The policy and legislation are not equipped enough to deal with all types of social networks threats which are increase day by day with more challenges, modern and sophisticated technologies.

d) Lack of tools and appropriate authentication mechanism to handle and deal with different security and privacy issues.

Because of the above mentioned factors that cause threats, we recommended the following strategies for circumventing threats associated with social website:

a) Building awareness the information disclosure: users most take care and very conscious regarding the revealing of their personal information in profiles in social websites.

b) Encouraging awareness -raising and educational campaigns: governments have to provide and offer educational classes about awareness -raising and security issues.

c) Modifying the existing legislation: existing legislation needs to be modified related to the new technology and new frauds and attacks.

d) Empowering the authentication: access control and authentication must be very strong so that cybercrimes done by hackers, spammers and other cybercriminals could be reduced as much as possible.

e) Using the most powerful antivirus tools: users must use the most powerful antivirus tools with regular updates and must keep the appropriate default setting, so that the antivirus tools could work more effectively.

f) Providing suitable security tools: here, we give recommendation to the security software providers and is that: they have to offers some special tools for users that enable them to remove their accounts and to manage and control the different privacy and security issues.

## 6. FUTURE TRENDS OF SOCIAL NETWORKING WEBSITES

In spite of the development and advanced technologies in social networking websites adjustment, a few are listed as below:

a) A need for more improvements for social networks so that they can allow users to manage their profiles and connecting tools.

b) A need for convergence and integration of social networks and future virtual worlds.

c) Needs for data integration from different networks, i.e. identification of all contents related to specific topic. This needs particular standards and sophisticated technology supported by social networks providers.

d) Many social networks need standard application programming interfaces, so that users can import and export their profiling information by using standard tools. (For example, Facebook and Google have applied new technologies that allow user data portability among social websites, representing a new source of competition among social networking service).

We hope that in the near future, one can by single sign-in functionality use over websites, that is, the user IDs are portable to other websites.

Moreover, virtual worlds have distinct virtual economies and currency that based on the exchange of virtual goods. Games are one of the newest and most popular online application types on social websites. Here, we have to mention the importance of privacy and security to save users from fraudsters who attempt to steal social networking credentials and online money.

Finally, we have to mention that the advances in the social websites and mobile-phone usage will effect on the growing of using mobile social networking by adding more features and application not only to mobiles, but also to social televisions for future chat, email, forums, and video conferencing [8, 9].

## 7. RISKS PREVENTION AND THREATS VULNERABILITIES

In this Section, we supply with some important recommendations to help social network users stay save by applying the followings:

a) Always have very strong passwords on your emails and other social web sites
b) Limiting the provided personal information in the social web sites as much as you can
c) Change your passwords regularly, so that your information can be out of reach by hackers.
d) Provide with the minimum amount of information to the website and internet due to the publicity of the internet.
e) Don't trust online others and don't answer on special questions from unknown users or companies i.e. be sceptical.
f) Check privacy policies and be aware of unknown emails and links provides by unknown users.
g) To prevent detecting emails address by spammer techniques, write the email: xyz@hotmail.com as xyz at hotmail dot com.

## 8. CONCLUSION

Although social networking websites offer advanced technology of interaction and communication, they also raise new challenges regarding privacy and security issues. In this paper, we briefly described the social networking web sites, summarized their taxonomy, and highlighted the crucial privacy and security issues giving some essential antithreats strategies with the perspective of the future of the social networking websites.

We think that the advancement of new technology in general and social websites in particular will bring new security risks that may present opportunities for malicious actors, key loggers, Trojan horses, phishing, spies, viruses and attackers. Information security professionals, government officials and other intelligence agencies must develop new tools that prevent and adapt to the future potential risks and threats. It can also safely manipulate the huge amount of information in the internet and in the social websites as well.

**Authors**

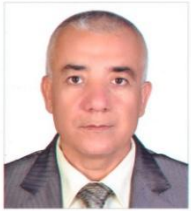

**Wajeb Gharibi** is an Associate Professor and Chairman, Department of Computer Engineering & Networks, College of Computer Science & Information Systems, Jazan University, Jazan, Kingdom of Saudi Arabia. He obtained his earned Ph.D. degree in Informatics from Institute of Mathematics and Computer Science, Byelorussian Academy of Sciences, Belarus in the year 1990. His research interest includes Information Security, Microelectronics, SoC, Design & Analysis of Network algorithms, Combinatorial Optimization, Operations Research and Data Analysis. He got many National and International Prizes and he has published more than 70 Research papers in National and International Journals and Conferences.

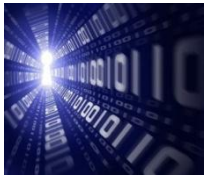

**Maha Shaabi** is a Teaching Assistant in the Department of Computer Science at the College of Computer Science & Information Systems, JAZAN University, Jazan, Kingdom of Saudi Arabia. She received her Bachelor degree in Information technology from King Saud University, Saudi Arabia in the year 2009. Her research interest includes Network security, Network algorithms and programming.